\documentclass[reprint,NumberedRefs]{JASAnew}

\begin{document}

\title[submitted J. Acoust. Soc. Am.]{Locating Objects with Signal Times Amongst Shadows and Black Holes in Two-Dimensional Models}


\author{John L. Spiesberger}


\affiliation{Dept. Earth and Environment Science, U. Pennsylvania, Philadelphia, PA 19104-6316, USA}


\email{john.spiesberger@gmail.com}
\thanks{}





\begin{abstract}
Calling mammals, ships, and many other objects have been commonly located during the last century
with two-dimensional
(2D) models from measurements of signal time even when the objects are not on the 2D surface.  
The overwhelmingly common method for locating signals with 2D models takes signal speed as constant.
Distance is computed by multiplying this speed
by signal time. For monostatic, bistatic, and Time Differences of Arrival (TDOA) measurements,
the distances constrain locations to  circles, ellipses, and hyperbolas respectively, whose intersections
yield location. However,
the speed needed to obtain correct locations depends on the horizontal separation between object and
instrument. In fact, if their horizontal separation is zero the speed needed for correct location must also be zero. In light
of this singularity, methods are derived for generating extremely reliable confidence intervals for location
and identifying  regions of the 2D model where a 3D model is needed. Because speeds needed for correct
location are spatially in-homogeneous in the extreme, isosigmachrons and isodiachrons emerge as natural
geometries for interpreting location instead of ellipses and hyperbolas.
These issues are caused by choice of coordinates,
and the same phenomena occur in general relativity regarding the speed of light
and black holes.

\end{abstract}


\maketitle


\section{\label{sec:1} Introduction}

Locations of a wide variety of objects and phenomena
are often estimated with a two-dimensional (2D) model from measurements of the propagation times
of signals. The objects and signal-measuring instruments are almost never on the 2D model's surface: e.g.
a Euclidean plane. In other words, the location of the object is explicitly forced by the 2D model to reside on the 2D surface
even though it is not there. There are tens of thousands of papers discussing these models dating to at least 1918 \cite{bateman}.
A Google search with ``TDOA 2D location'' yields 69000 sites, where TDOA stands for Time Difference of Arrival.
Modeling locations in 2D is ubiquitous. Contemporary examples include locating calling mammals in the ocean
\cite{ishmael,sbe,spermwhale,warner_and_dosso}, sounds in a room via robots \citep{6038205}, ships \citep{suwal}, cell phones \citep{cell_phones},
  lightning \citep{lewis}, wildlife radio transmitters \citep{kruger},
aircraft radio emissions \citep{lee_lee}, bistatic sonars \citep{craparo}, and theoretical developments \cite{jin}.
These models derive locations by translating TDOA or bistatic signal times to difference or sum of distance assuming signal speed is
constant \cite{ishmael,spermwhale,jin,cell_phones,kruger,lee_lee,suwal,lewis,craparo,6038205},
the overwhelmingly-common
case, or are constrained to a finite interval \cite{sbe,warner_and_dosso}, e.g. [1450 , 1500] m/s for the ocean.
Over the last century, it appears all authors  missed  a fundamental problem. This oversight
appears to call into question some results from some studies.

The problem is illuminated by calculating the signal speed needed for
a correct location when the object is near a receiver but not on the 2D surface. Suppose an
object has the same horizontal coordinate as this receiver but is  100 meters above. Its signal propagates at 1 m/s so arrives 100 s
after emission. In the 2D model, the object is zero meters from the receiver, so the
speed to use to get the object's horizontal location to equal the receiver's location is zero meters divided by 100 s: zero meters
per second!  We cannot find any previous reference for this fact and is the raison d'etre for this study. 
In this light, we quantify the regions of validity of 2D models, show how to extend their validity,
provide a method for deriving extremely reliable confidence intervals for location, and explain how unconventional geometrical shapes
naturally emerge as a means to derive reliable locations.  The topics are related to coordinate systems in general,
with connections to other fields (Sec. VII).  Most of the ideas in this paper appear to be new, with new elements 
noted where needed.

We use three phrases to describe signal times used for location. They are,
{\it Direct-path time}: a measurement of time between two objects such as a source and receiver, 
or from a transceiver whose signal reflects
from an object. In the latter case, we divide time by two to get the equivalent time from transceiver to object.
{\it TDOA}: the difference of signal propagation time between a source and two receivers.
{\it Bistatic-time}: time of travel from source to reflector to receiver not co-located with source. 

The singularity does not appear when the coordinate system is changed from 2D to three-dimensions (3D).  Usually, but not always,
scientists intuitively use 3D models when objects are far from
a 2D modeling plane.  We show, perhaps for the first time,
how to draw regions within a 2D model-surface where the 2D approximation is invalid and where 3D models are needed
for reliable locations.
There are many ways of modeling
locations in 2D and 3D, and the best choice is application-dependent.  Li {\it et al.} \citep{li_etal} review 2D and 3D modeling
with direct-path times and TDOA, Cummins and Murphy \citep{cummins_murphy} review 2D and 3D models to locate lightning,
Rascon and Meza review means to locate sounds with robots \citep{rascon}.

Reliable locations are important for at least two reasons.
Firstly, we wish to understand the behavior of a vocalizing marine mammal in the presence of 
disturbing sounds such as air-guns or Navy sonars. Reliable locations are needed for censusing and understanding 
behavior.  Secondly, We wish to track locations of sounds.
If the model for location yields incorrect locations because the 2D approximation is invalid, the tracker
is provided invalid data. Valid locations have a better chance of forming valid tracks.

It is natural to ask why 2D models are used when it is widely known they are approximations. Reasons include,
objects are not usually near receivers,
3D models require too much computer time,
3D models are more complicated,
2D models are ubiquitous, implying their validity,
receivers are all near the same vertical coordinate and cannot be used to estimate vertical coordinates of a source
and,  we are happy with the object's horizontal location: there is no need for a
3D model.

The subject of this paper may seem disorienting to readers familiar with 3D models but less so with 2D.
For 3D models in the ocean and atmosphere, the speed of sound can vary vertically and horizontally by significant
amounts. From this perspective, some readers may wonder why any contemporary paper would discuss 2D models where speed is
either constant, or, in a few cases, varies between specified bounds. For the reasons stated above, 2D models
are commonly used and understanding how to use them appropriately is interesting and important in some
circumstances.

We explain the problem from the perspective of planar 2D models, i.e. Flatland, (Sec. II) and quantify errors when
the 2D effective speed is constant (Sec. III).  The material in these two sections appears to be new to science.
Errors are eliminated when
this speed  is spatially inhomogeneous with location interpreted with unconventional geometries
(Sec. IV). The unconventional geometries are not new, but this appears to be the first
publication to explain how their use eliminates errors from the 2D approximation.
Sec. V exhibits a method yielding extremely reliable confidence intervals in 2D models, accounting
for all relevant contributions to error, including measurements of signal times, instrument locations,
and 3D effective speeds.  This section shows how to identify invalid regions of 2D models, in which
3D models must be used to obtain reliable locations.  Showing how to identify invalid regions appears
to be new.
Sec. VI discuses other flatlands, namely spherical and spheroidal surfaces. This
material appears to be new.
Results are discussed in Sec. VII and connections are made with other fields.

\section{\label{sec:2} Singularities in Flatland}

\subsection{\label{subsec:2:1} Two- and Three-Dimensional Effective Speeds}

\begin{figure}[ht]
\includegraphics[width=\reprintcolumnwidth]{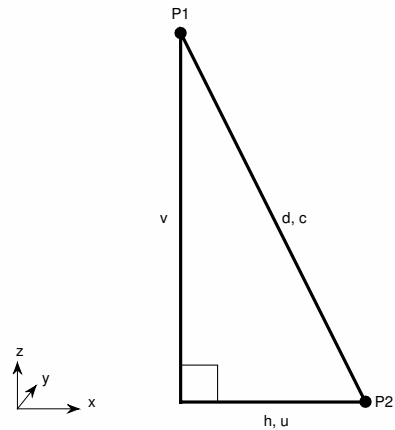}
\caption{\label{fig:triangle_defn}{Signal propagates between points P1 and P2. Distance of separation is d, with horizontal and vertical
separations h and v. xy-plane is defined to be ``horizontal'' where locations are obtained 
from model. Both points may be out of xy-plane. ``Effective 3D and 2D speeds of signal are c and u respectively where
horizontal separation, h, is parallel to xy-plane.}}

\raggedright
\end{figure}

Suppose we wish to model locations of objects emitting or reflecting signals.
A signal propagates between points P1 and P2 in 3D space (Fig.~\ref{fig:triangle_defn}). A measurement
is made at  P2. The signal propagates between the points following the laws of physics, not usually the
line segment of length, $d$, unless the speed of the signal is spatially homogeneous.

Define the 3D effective speed, $c$, to be the geodesic distance between the points, $d$, divided by the time, $t$, 
for the signal to propagate in between,
\begin{equation}
c \equiv d/t \ . \label{eq:3d_effective_speed}
\end{equation}
In flat space, the geodesic length is the Euclidean distance.
The 2D effective speed, u, is adopted by the 2D model for location. It is defined to be the horizontal separation, $h$,
between the points divided by the same propagation time, $t$,
\begin{equation}
u \equiv h/t \ . \label{eq:2d_effective_speed}
\end{equation}
Solve  Eq. (\ref{eq:3d_effective_speed}) for $t$ and substitute into Eq. (\ref{eq:2d_effective_speed}) to get,
\begin{equation}
u = \frac{hc}{d} \ . \label{eq:2d_speed1}
\end{equation}
2D and 3D effective speeds are identical when $h=d$: both points are on the 2D surface.

To see how the 2D effective speed depends on horizontal and vertical separation instead of horizontal and 3D separation, 
we use the Pythagorean relation, $d=(v^2+h^2)^{1/2}$,  for $d$ in Eq. (\ref{eq:2d_speed1}),
\begin{equation}
u = \frac{c}{(1+(v^2/h^2))^{1/2}} \ . \label{eq:2d_speed2}
\end{equation}
When either P1 or P2 are  not on the 2D surface, $v$ is not zero,  and the 2D and 3D effective speeds differ.
When the horizontal separation is zero and the vertical separation is not zero, the denominator in Eq. (\ref{eq:2d_speed2}) 
goes to infinity and the  effective speed is {\it zero}. 
Eq. (\ref{eq:2d_speed1}) shows the same behavior: the length of the hypotenuse, $d$, exceeds the horizontal separation, $h$, when
the vertical separation is positive, so
when $h$ goes to zero, $d$ remains positive and $u$ goes to zero. The zeros of effective speed are a problem for 2D models: they
are singularities of the approximation caused by removal of the third spatial dimension.

2D and 3D effective speeds are not the same  as ``effective speed in a moving medium'' the
sum of the scalar speed of sound with the speed of advection of the medium.

\subsection{\label{subsec:2:2} Locating Signals in Flatland}

Pretend we live on Flatland and know nothing of
3D space. 
Flatland transmits music via ``radio'' 
whose transmissions are delayed by a signal speed equal to 1450 m/s.
Some listeners do not receive the signal so scientists investigate.  
They hypothesize the existence of a reflecting object: it scatters signals so the combination of
the direct and scattered signals destructively interfere. To find the reflector, they build an instrument
to transmit and receive pulses: a ``radar.'' They measure the distance to a reflecting object by measuring
the round-trip travel time of the signal:  $T$.  If they measure time $T_1$, they
reason the object is at distance $l_1=c/(2T_1)$ where $c$ is signal speed.   
The first measurement, $T_1$, yields distance $l_1$, so the object is on a circle of
radius $l_1$.  They move the radar, 
make a second measurement wherein $T_2$ yields a distance $l_2$ whose corresponding circle
intersects the first at two points. A third measurement, $T_3$ yields
a third circle intersecting one point.
They drive to the hypothesized location and
find  a metal dog house whose reflected signal
cancels the direct path. The radio station pays for a non-reflecting wooden
dog house, the metal is scrapped, and the problem is solved.

\begin{figure}[ht]
\includegraphics[width=\reprintcolumnwidth]{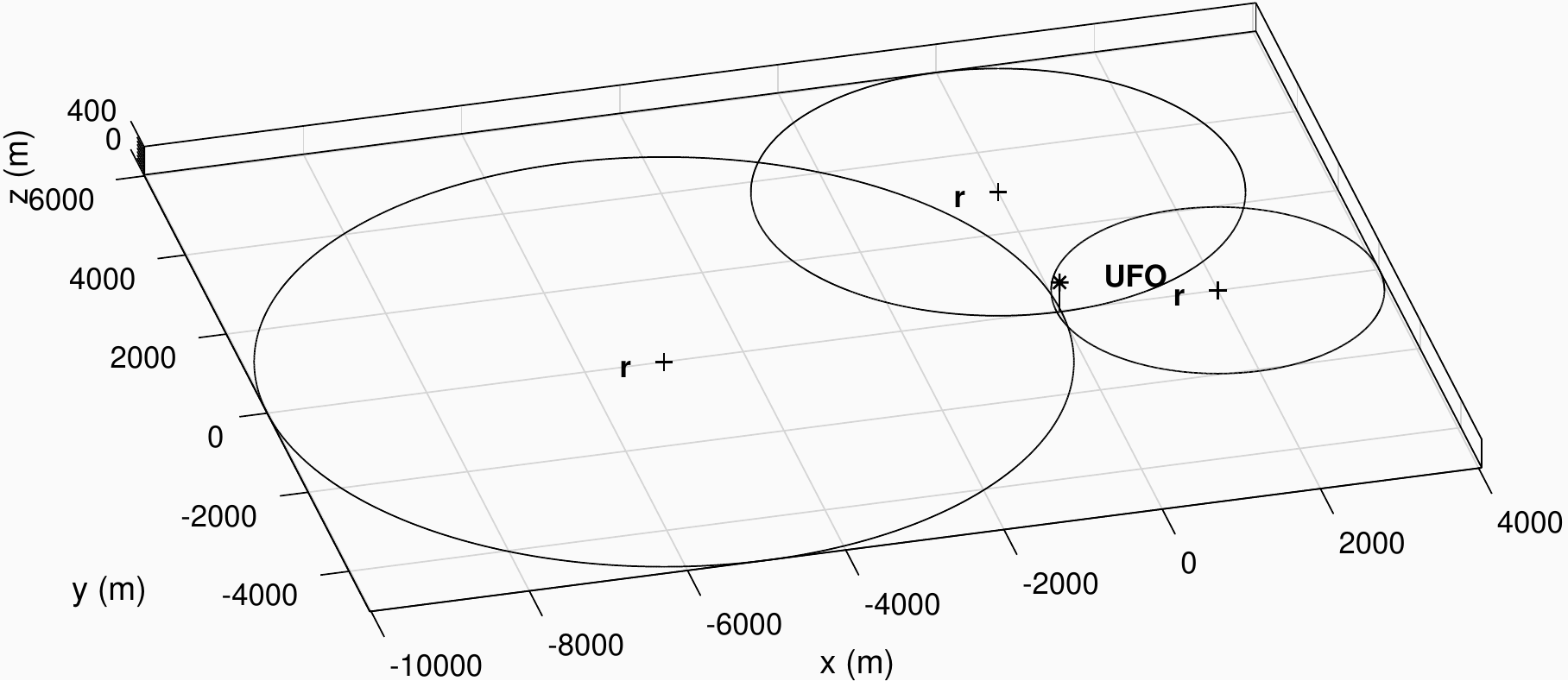}
\caption{\label{fig:ufo_far_from_radars}{UFO (*) parked over Flatland. Each of three radars (r +) yield UFO's location
on circle {\it in} Flatland since they know nothing of 3D space (Sec. IIB).
Circles intersect within 400 m of UFO's horizontal location (Fig.~\ref{fig:ufo_far_from_radars_birds_eye_view}).}}

\raggedright
\end{figure}

\begin{figure}[ht]
\includegraphics[width=\reprintcolumnwidth]{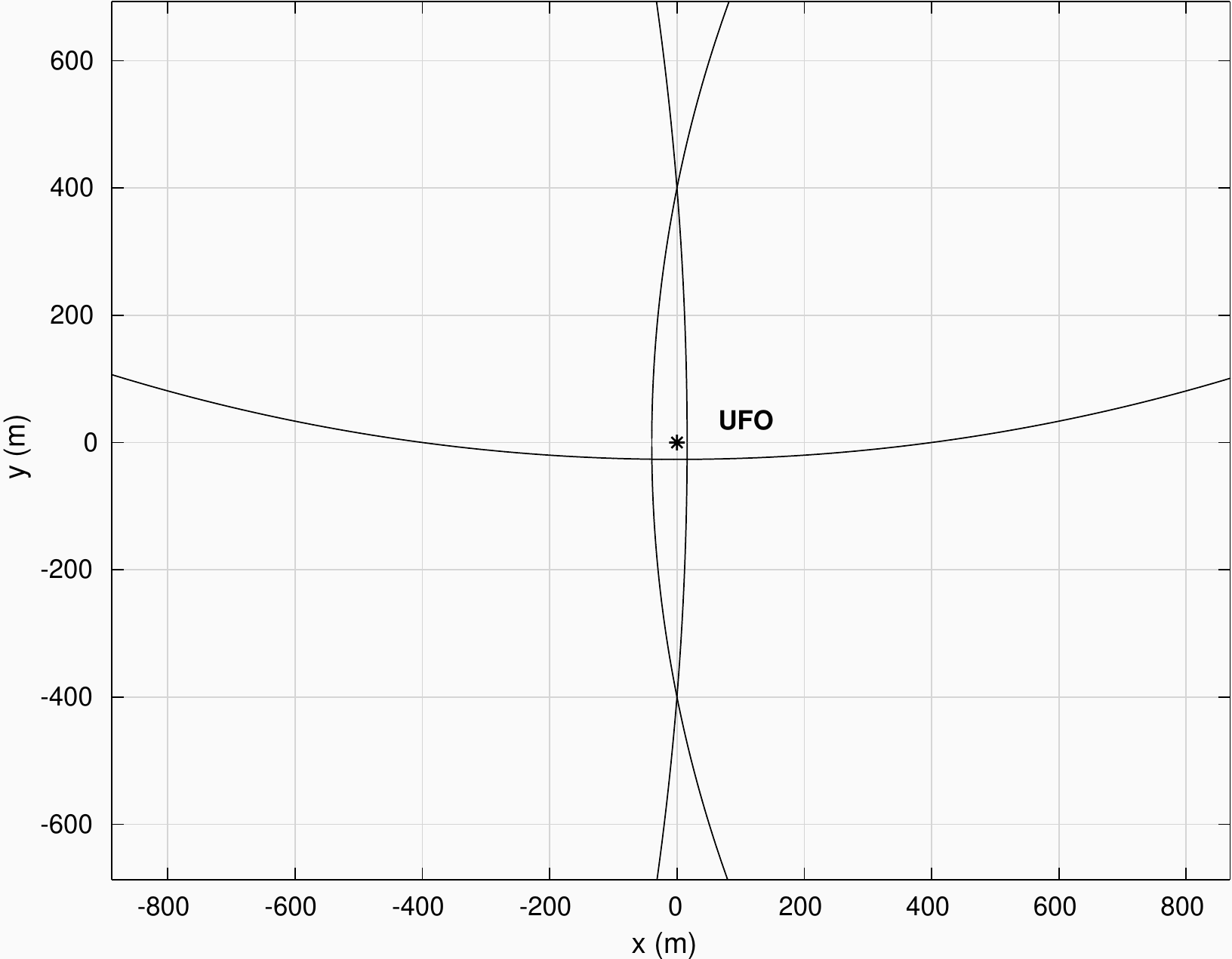}
\caption{\label{fig:ufo_far_from_radars_birds_eye_view}{Same as Fig.~\ref{fig:ufo_far_from_radars} except viewed above
    showing circles not intersecting at single point.}}

\raggedright
\end{figure}

A few years later, other families find their radios no longer pick up broadcasts.  
Scientists re-deploy their radar, unaware of the existence of an Unidentified 
Flying Object (UFO) in 3D space parked over Flatland with Flatland Cartesian location (0,0). 
They deploy radars at three locations, and obtain three circles intersecting 
in the proximity of the origin (Fig.~\ref{fig:ufo_far_from_radars}). From
afar, the intersections look like they yield a useful solution, but when they enlarge their figure
the points of intersection differ by hundreds of meters (Fig.~\ref{fig:ufo_far_from_radars_birds_eye_view}).
They are unsatisfied because the accuracy of their measurements should yield a single location.
Furthermore, they drive to the area containing the points and find an open field incapable of
scattering signals.
They know the physics of the problem very well: signal speed 
is known to nine significant digits (1450.00000 m/s) as are the locations of their radars and measurements of 
signal time. 

\begin{figure}[ht]
\includegraphics[width=\reprintcolumnwidth]{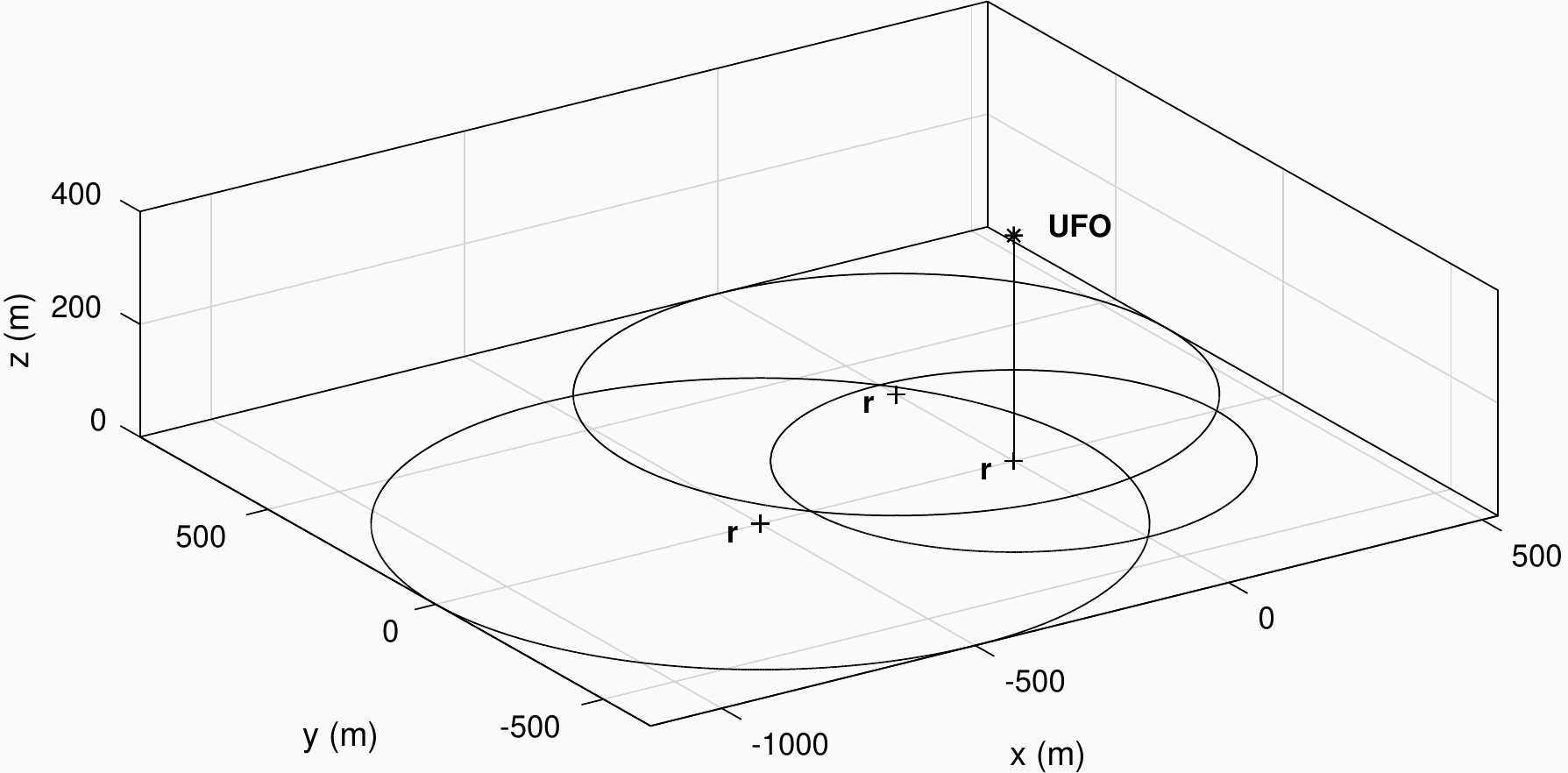}
\caption{\label{fig:ufo_near_radars}{Same as Fig. \ref{fig:ufo_far_from_radars} except Flatland scientists move radars close
to origin, hoping for more accuracy. Three new circles do not come close to yielding
a single point of intersection: things are worse.}}

\raggedright
\end{figure}

Because Flatland scientists cannot explain 400-m discrepancies in location, they move
their radars close to origin hoping for more accuracy, but instead
obtain worse results
(Fig. \ref{fig:ufo_near_radars}). In response, their theoretical physicists
hypothesize the existence of a third spatial dimension of the universe, invent
a new geometrical shape called a ``sphere'', use the same data to intersect three spheres, whose
intersection does indeed occur at a single point: their $x-y$ origin and elevation $z=400$ m: the
true location of the UFO.

\section{\label{sec:3} Quantifying errors with constant 2D effective speed}

A common method for locating objects in 2D models is to assume  the 2D effective speed is a constant.
We quantify errors of this approximation, leaving discussion of  other errors affecting
location to  Sec.  V.

\subsection{\label{subsec:3:1}  Direct-Path Time}

Suppose we estimate the time, $t$, for a signal to propagate between a source on Flatland and a
reflector at perpendicular distance, $v$, from the surface (Fig. \ref{fig:triangle_defn}).
Flatland corresponds to $z=0$ in a Cartesian coordinate system.  Let the source be located
at $x=0$ and $y=0$.  The reflector is located at $(x,y,z=v)$. In Flatland, the distance
to the reflector is 
\begin{equation}
d_{flat}=c_{flat}t \ , \label{eq:d_flat}
\end{equation}
where $c_{flat}$ is the single modeled signal speed speed in Flatland. In
3D space, the 3D distance to the signal is $d=ct$ where $c$ is the 3D effective speed 
(Fig. \ref{fig:triangle_defn}). The projection of $d$ onto Flatland is $h=(d^2-v^2)^{1/2}$, or
\begin{equation}
h=d(1-v^2/d^2)^{1/2}  , \label{eq:h}
\end{equation}
(Fig. \ref{fig:triangle_defn}). The error of the 2D location model is,
\begin{equation}
\epsilon \equiv d_{flat}-h = c_{flat}t - h = c_{flat}d/c - h . \label{eq:epsilon}
\end{equation}
Substituting  Eq. (\ref{eq:h}) in Eq. (\ref{eq:epsilon}), we get,
\begin{equation}
\epsilon=\bigl[ c_{flat}/c - (1-v^2/d^2)^{1/2} \bigr] d . \label{eq:epsilon1}
\end{equation}

\begin{figure}[ht]
\includegraphics[width=\reprintcolumnwidth]{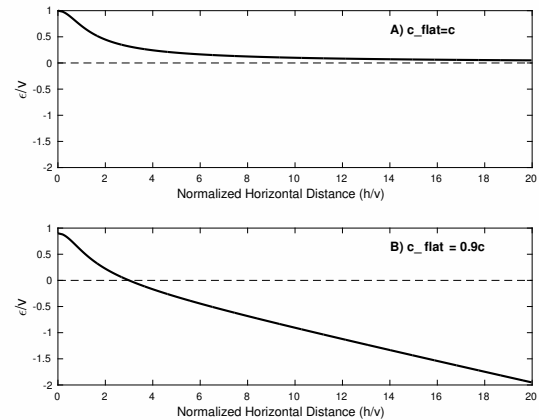}
\caption{\label{fig:norm_err_monostatic}{{\bf a)} Error of 2D location model versus true horizontal location of 3D object when
2D model uses 3D effective
signal speed, $c$. Horizontal distance to object projected onto 2D model normalized by perpendicular distance, $v$,
of object from 2D model (horizontal axis). Vertical axis is normalized error of 2D model,
$\epsilon/v$, where $\epsilon$ (Eq. (\ref{eq:epsilon1})). {\bf b)}: Same except 2D model uses nine-tenths
of 3D effective signal speed for purpose of decreasing 2D model errors at smaller horizontal distances at
expense of larger errors at large distances.}}

\raggedright
\end{figure}

A single speed is often adopted for the 2D model: the same as the 3D effective speed, $c_{flat}=c$.
This forces the error, $\epsilon$, to zero when the reflector's horizontal distance is much greater than
its vertical offset (Fig. \ref{fig:norm_err_monostatic}a). Distances are normalized
by vertical offset, $v$, because $v$ is the geometrical parameter affecting error. Because
errors are large when the horizontal offset, $h$, is small, we could choose a smaller value for $c_{flat}$
yielding smaller errors at small offsets and larger errors at large offsets 
(Fig. \ref{fig:norm_err_monostatic}b).

\subsection{\label{subsec:3:2}  TDOA}

We measure TDOAs from a source and estimate location with a 2D model. In some regions of the plane,
three receivers are sufficient to yield a unique mathematical solution, yet other regions
require four receivers \citep{schmidt}.  Let the signal time between the source and receiver $i$ be $t_i$.
With $R$ receivers, we measure TDOAs,
\begin{equation}
\tau_{ij} \equiv t_i - t_j \ , i < j; \ j= 2, 3, \cdots R. \label{eq:tau_ij}
\end{equation}
With three receivers, we measure
$\tau_{12}$ , $\tau_{13}$, and $\tau_{23}$ but without errors of measurement $\tau_{23}$ provides
no independent information since $\tau_{23} = -\tau_{12} + \tau_{13}$. Similarly,
four receivers yields three independent TDOAs.

We adopt a single sound speed in the 2D model, $c_{flat}$.  TDOAs are converted to a difference in distance 
from receivers 1 and $j$ with
\begin{equation}
l_1 - l_j = c_{flat} (t_1-t_j) = c_{flat}\tau_{1j} \ . \label{eq:diff_dist_hyp}
\end{equation}
This defines a hyperbola. Location can be obtained by intersecting hyperbolas.

We set the 2D effective speed as $c_{flat}=1450$ m/s: the same as the 3D effective speed. Consider a shallow-water
scenario with source at 15 m depth and four receivers at 50 m depth. We use four receivers to avoid the
mathematical plurality of solutions with only three receivers \citep{schmidt}. We assume $t_1$, $t_2$, $t_3$, and $t_4$
are measured without error, yielding three independent TDOAs. Two hyperbolas are intersected, each derived from 
$\tau_{12}$ and $\tau_{13}$. This yields  the ``first'' set of intersected locations in the plane with
0, 1, 2, 3, or 4 points of intersection.  If the source
was in the plane of the model, a solution would always exist, but not necessarily when the source
is out of the plane.  If there are two or more points
of intersection, data from the fourth receiver are used to resolve ambiguous locations.
We intersect the two hyperbolas associated with $\tau_{12}$ and $\tau_{14}$. These intersect
at the ``second'' set of locations containing 0, 1, 2, 3, or 4 elements.  If either the first or second set
is empty, no solution for location is determined.  If the first set contains two or more locations, and the second
set is empty, we end up with ambiguous solutions. Otherwise we choose the single location from the first set whose
distance is minimum to any of the locations from the second set.  If the source was in the plane of the model,
we would always have a single solution for location. The out-of-plane geometry introduces unavoidable
complications as long as we insist on using a 2D model with a single speed, $c_{flat}$.

\begin{figure}[ht]
\includegraphics[width=\reprintcolumnwidth]{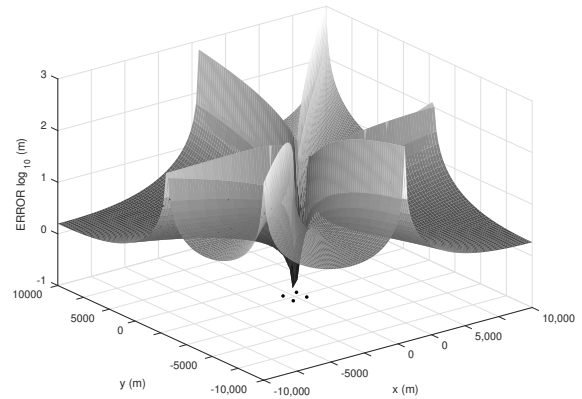}
\caption{\label{fig:model_dist_errors1_4rec_tdoa}{Maximum horizontal error of 2D (planar) location model
derived with TDOAs from four
receivers (dots) as function of horizontal location
of acoustic source. Source and receivers at 15 and 50 m depth respectively. Vertical axis is $\log_{10}$ of error: two is 100 m.
Errors due to non-coplanar objects and utilization of single 2D effective speed (Sec. III B).}
\label{f:model_dist_errors1_4rec_tdoa}}

\raggedright
\end{figure}

Receivers are placed at horizontal Cartesian coordinates (-510,500), (500,-490), (500,507), and (-502,506) m 
(Fig.~\ref{fig:model_dist_errors1_4rec_tdoa}). The source is placed at 200 m increments of $x$ and $y$
in an area $20 \times 20 \ \mbox{km}^2$ 
centered on the mean horizontal location of the receivers. The 2D model yields source locations $(x_m,y_m)$.
The error of each $(x_m,y_m)$ is its distance to the true horizontal location.
Unfortunately, very large errors occur at sub-grid intervals, so the plot greatly underestimates errors. It is impractical
to search the horizontal space with enough resolution to reveal the largest error.  For example, we decreased the grid interval
from 200 to 0.2 m near receiver 1. The maximum error rose to several hundred meters. Then the grid interval was
decreased to 0.1 m and the maximum error increased to 1781 m. Larger errors would likely be found with a finer mesh.

Errors are tabulated in five distance intervals from the mean location of the receivers (Table \ref{tab:table1}). Mean errors
are about 20 m. Maximum errors are large: between 600 and 1000 m.  When the source is located within the x-y perimeter
of the receiver's polygon, maximum error is at least 1781 m.
Errors are large compared with the mere 35 m offset of the source from the model plane.

\begin{table}[ht]
\caption{\label{tab:table1}2D model error in Fig. \ref{fig:model_dist_errors1_4rec_tdoa} due only to source 
being out of 2D  model plane. Distances computed with respect to mean horizontal location of receivers. 
Maximum horizontal distance of any receiver from mean is 716 m: Corresponding minimum, mean, and maximum errors 
in interval [0 , 716] m are 0.14, 1.3,  and at least 1781 m respectively. Maximum error difficult
to compute (Sec. III B)}
\centering
\begin{tabular}{rrrr}
\hline
\multicolumn{1}{l}{DISTANCE}&
 \multicolumn{3}{c}{2D MODEL ERRORS (m)}\\
 \multicolumn{1}{l}{INTERVAL (m)}&
 \multicolumn{1}{c}{MINIMUM}&
  \multicolumn{1}{c}{MEAN}&
   \multicolumn{1}{c}{MAXIMUM}\\
\hline 
0 to 1999& 0.14& 16& $\geq$1781  \\
1999 to 3999& 1& 18& $\geq$348 \\
3000 to 5998& 1& 18& $\geq$568\\
5998 to 7998& 1& 19& $\geq$726\\
7999 to 9997& 1& 19& $\geq$831 \\
\end{tabular}
\end{table}

\subsection{\label{subsec:3:3}  Bistatic time}

\begin{figure}[ht]
\includegraphics[width=\reprintcolumnwidth]{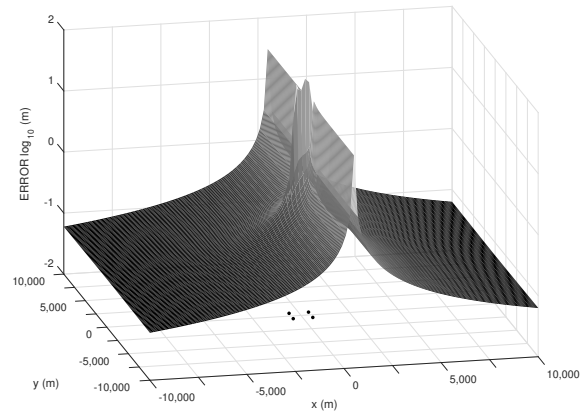}
\caption{\label{fig:model_dist_errors_bistatic_run4}{Same as Fig. \ref{fig:model_dist_errors1_4rec_tdoa} except
for bistatic-times where four dots comprise one source and three receivers (Sec. IIIC).
Locations derived by intersecting ellipses
from perfect measurements. Errors
caused by target lying out of model plane.}}

\raggedright
\end{figure}

We locate a target by measuring the bistatic time for a signal to leave a  source at 50 m depth, reflect from
a target at 15 m depth, and be received at three receivers at 50 m depth.
Locations are computed
by intersecting ellipses because an ellipse is the locus of points whose sum of distances
is constant.  Everything else is the same as Sec. B.
The target would be located perfectly if it was in the same plane as the instruments.
Errors are generated with a grid interval of 200 m yielding 82 m near the instruments
to 30 m far away (Fig. \ref{fig:model_dist_errors_bistatic_run4}).    We searched with a 
fine grid near transceiver one, but found no large errors.  We do not know if a finer grid would
yield larger errors as was found for hyperbolas (Sec. IIIB). Unlike hyperbolas extending to infinity,
ellipses have finite extent and impose an upper error limit.

\section{\label{sec:4} Eliminating Errors in Flatland with inhomogeneous 2D effective speed and unconventional geometry}

The previous section quantified errors from  2D models when the 2D effective speed is constant.
None of this matters unless  we want a reliable confidence interval for location: the subject of
this section.  Without loss of generality, we assume  2D locations are on a horizontal $x-y$ plane
and the object to be located has Cartesian coordinates $(x,y,z)$.  We could of course side-step all these problems using
a 3D location model. We consider four other remedies (Sec. 4.1, 4.2, 4.3, and 4.4).

\subsection{\label{subsec:4:1}   2D Model in Valid Regions}

First, errors of  2D location are quantified for a specified choice for the 2D effective speed using methods described in Sec. III. 
Let this horizontal error be $E(x,y)$.  Let $\hat{E}$ denote the maximum acceptable error.  We receive $I$ signal-time data, and 
compute their locations with the 2D model: $(x_i,y_i) \ , \ i=1,2,3, \cdots I$. We accept the $i'th$  location when $E(x_i,y_i)<=\hat{E}$. 
Otherwise the datum is discarded.  In this scenario, there remain holes in the 2D model where locations are never estimated.

\subsection{\label{subsec:4:2} Effective Speed is Function of Measured Signal Time}

The idea is to improve the accuracy of locating a signal by letting the 2D effective speed be
a function of the measured signal time(s). Let $U(t,{\bf r}_j)$ approximate
the 2D effective speed as a function of the measured signal time,  $t$, and instrument location(s), ${\bf r}_j$.
Explicit dependence of $t$ on the object's location is implied but not shown.
We think of $U(t,{\bf r}_j)$ as a single value, i.e. a 2D effective speed, or more generally a confidence
interval or some statistical summary of the 2D effective speed for each horizontal location in the model plane. 
Possible statistical summaries include mean and median values.
A procedure for constructing $U(t,{\bf r}_j)$ is
1) Specify locations of the instruments, ${\bf r}_j$,
2) Specify a 3D grid of locations, $(x_k,y_k,z_k)$  where the 2D effective speeds are computed at the instruments,
3) Specify the 3D effective speed, $c$,
4) Use Eq. (\ref{eq:2d_speed2}) to compute the 2D effective speed, $u(x_k,y_k,z_k)$, for each location in the grid, and
5) Compute $U(t,{\bf r}_j)$ from $u(x_k,y_k,z_k)$.

Here is an example. We want a 100\% confidence interval for the 2D effective speed between instrument one and
a fixed horizontal location $(x_k,y_k)=(X,Y)$ in the 2D model plane.  $Q$ values of 
$k$ yield the same coordinate, $(X,Y)$, but with different vertical coordinates, $z_k$.
The minimum 2D effective speed at this point, $\check{u}$, is
the minimum of $u(x_k,y_k,z_k)$ among all $Q$ vertical coordinates,
$z_q \ ; q=1,2,3, \cdots , Q$.
The maximum, $\hat{u}$, is obtained similarly.  The desired confidence interval is
$U(t,{\bf r}_j)=[\check{u} , \hat{u}]$. The symbols $\check{\mbox{}}$ and $\hat{\mbox{}}$ indicate minimum and 
maximum values respectively.

\subsubsection{\label{subsec:4:2:1} Direct-path times}

For direct-path times, we receive a measurement of signal time, then use $U(t,{\bf r}_1)$ to obtain an 
estimate of the 2D effective speed.
If we desire 100\% confidence intervals for horizontal location, we compute 
$U(t,{\bf r}_j)=[\check{u} , \hat{u}]$, and draw
the  annulus about ${\bf r}_1$ whose inner and outer radii are $\check{u} t$ and  $\hat{u} t$ respectively.
The procedure is repeated for the signal time measured at a second instrument, yielding a second annulus. 
The object's horizontal
location resides in the  intersection of the two annuli: either one region or  two non-overlapping regions. 
Data from a third instrument yields a third annulus whose intersection yields one or two contiguous regions 
of the plane.
If we desire locations with less than 100\% confidence, we repeat the procedure using p\% confidence intervals for $U(t,{\bf r}_j)$.
For example if we choose $p=95\%$, then each annulus has probability 0.95 of containing the true location of the signal.
Each annulus is statistically independent, so the intersection of three annuli has probability equal to $p^3$ of containing
the signal's horizontal location. If we want the final region to be valid with a probability of P percent we choose
$p=P^{1/3}$.  A geometrical interpretation of location is  made with the picture of 
intersecting
circles or annuli, the projections  of spheres  or thick spheres onto a horizontal plane.

\subsubsection{\label{subsec:4:2:2} TDOA}

For TDOA data, the procedure for estimating $U(t,{\bf r}_j)$ is the same as data for direct-path times. 
However, when it comes to 
locating signals in the 2D plane, the problem is different: we will see location cannot be interpreted
by intersecting hyperbolas.  When the 2D effective
speed is not the same value for each section, we have the problem of finding the locus of points in space satisfying,
\begin{equation}
t_1 - t_2 = \frac{l_1}{c1_{flat}} - \frac{l_2}{c2_{flat}}   \ , \label{eq:tdoa_isodiachron}
\end{equation}
Here, the 2D effective speeds between the object and receivers 1 and 2 are $c1_{flat}$ and $c2_{flat}$ respectively.
If they are equal, Eq. (\ref{eq:tdoa_isodiachron}) becomes,
\begin{equation}
t_1 - t_2 = \frac{l_1 - l_2}{c1_{flat}}   \ , \label{eq:tdoa_hyperbola}
\end{equation}
and multiplying by $c1_{flat}$ yields Eq. (\ref{eq:diff_dist_hyp}) defining the hyperbola: the locus of points 
in space whose difference in distance is constant from two points.

\begin{figure}[ht]
\includegraphics[width=\reprintcolumnwidth]{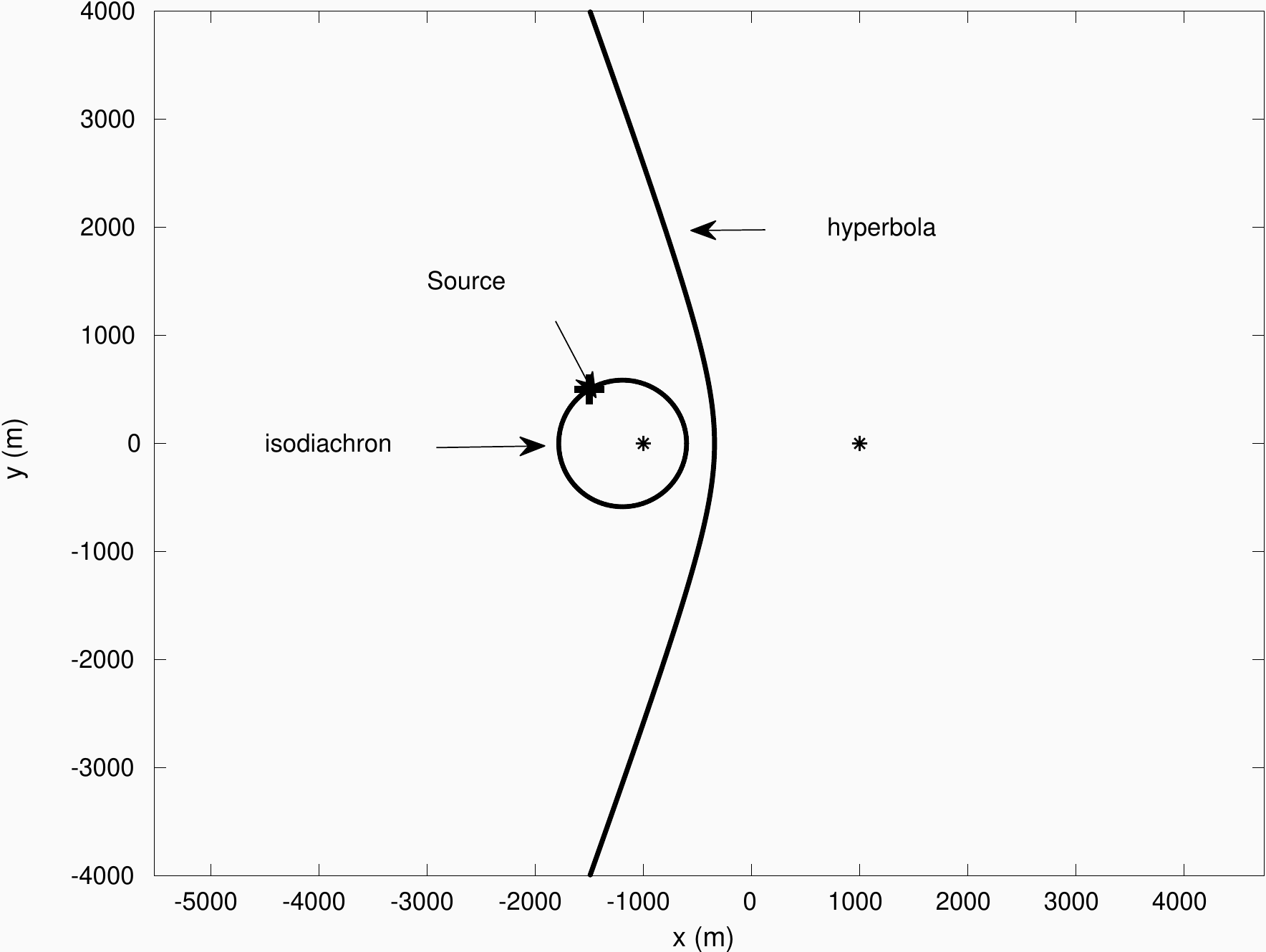}
\caption{\label{fig:isodia}{Geometrical shapes for locating horizontal location of object in ocean with a 2D model
from measurements of Time Differences of Arrival at 
two receivers (asterisks) assuming 2D effective speed is spatially homogeneous (hyperbola)
and inhomogeneous (isodiachron).  Receivers about 4000 m deeper than source (Sec.  IVB2). Hyperbolas are
inappropriate for locating signals. True horizontal location of source is on isodiachron (+).}}

\raggedright
\end{figure}

The locus of points in space whose difference in {\it signal time} is constant is an isodiachron, derived from the Greek
words iso for same and diachron for difference in time \citep{isodiachrons}.  This is a  natural geometry for understanding
locations when the effective speeds differ, as for this 2D location model. 
Consider its shape in the ocean where two receivers are on the bottom
at depth 4000 m and on the $x$ axis at $\pm 1000$ m. The source is nearby horizontally, (-1500, 500), but not vertically,
because we set its depth to 15 m.
The 3D effective speed is
$c=1450$ m/s and the 2D effective speeds are derived with Eq. (\ref{eq:2d_speed2}): $c1_{flat}=253.33$ m/s and  $c2_{flat}=781.43$ m/s
respectively.  The measured TDOA is 2.7912 - 3.2626 = -0.4714 s. The isodiachron looks like a circle for this case,
and it intersects the true location of the source (Fig. \ref{fig:isodia}). The hyperbola is drawn for a difference of distance
given by $c (t_1-t_2) = 1450(-0.4714) \mbox{m} =-683.5$. It does {\it not} contain the true location (Fig. \ref{fig:isodia}).  
Isodiachrons do not always look
like circles: sometimes they look ellipse-like and other times even more convoluted (i.e. Fig. 1b from \citep{isodiachrons}).
Unlike hyperbolas, they never extend to infinity when the effective speeds differ: 
a desirable quality of any geometrical interpretation of location \citep{isodiachrons}.

Confidence limits for direct-path times were annuli (Sec. IV, B1). Isodiachrons do not not maintain the same
shape as effective speeds change: we cannot zoom them in and out as we did for circles to get confidence intervals.
Instead, we could choose many pairs of 2D effective speeds within the desired confidence interval, and draw the isodiachrons
for each realization: they will fill a finite region of space, and the realizations can be plotted to show
corresponding confidence intervals. Alternatively, confidence intervals could be computed with
Sequential Bound Estimation \citep{sbe,patent_sbe1},  a technique discussed later.

\subsubsection{\label{subsec:4:2:3} Bistatic-times}

For bistatic data, estimating $U(t,{\bf r}_j)$ is the same as with TDOA.
However, locations cannot be interpreted
by intersecting ellipses.  Since the 2D effective
speed differs for each section, we need to find the locus of points satisfying,
\begin{equation}
t_1 + t_2 = \frac{l_1}{c1_{flat}} + \frac{l_2}{c2_{flat}}   \ . \label{eq:time_isosigma}
\end{equation}
If $c1_{flat}=c2_{flat}$, Eq. (\ref{eq:time_isosigma}) is,
\begin{equation}
t_1 + t_2 = \frac{l_1 + l_2}{c1_{flat}}   \ . \label{eq:time_ellipse}
\end{equation}
Multiplying by $c1_{flat}$ yields a definition of the ellipse:  the locus of points 
in space whose sum in distance is constant from two points.

\begin{figure}[ht]
\includegraphics[width=\reprintcolumnwidth]{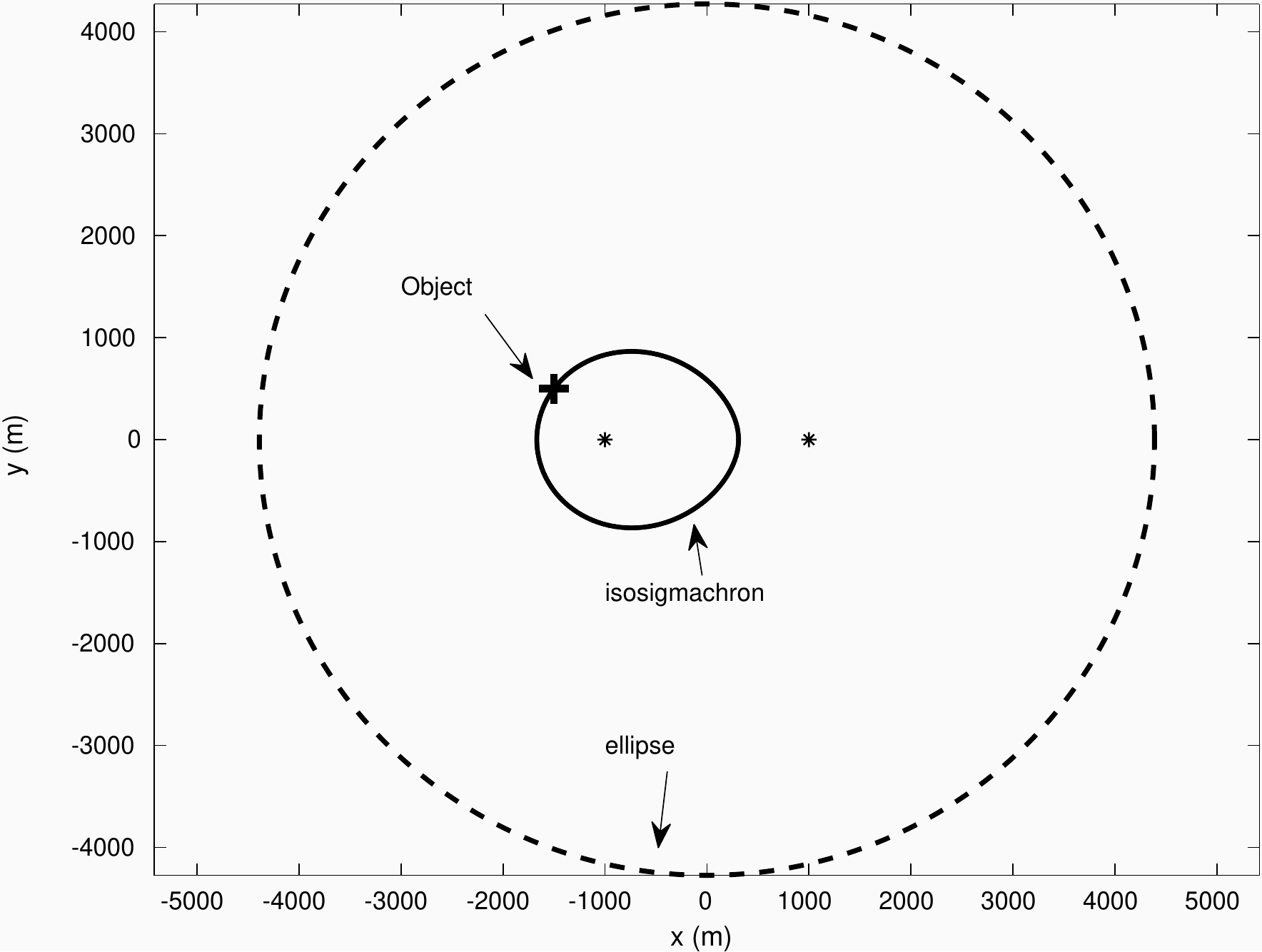}
\caption{\label{fig:isosigma}{Geometrical shapes for locating horizontal location of reflector in ocean with 2D model
from bistatic signal time assuming 2D effective speed is spatially homogeneous (ellipse)
and inhomogeneous (isosigmachron).  Source (*) and receiver (*) about 4000 m deeper than reflector (Sec. IV B3). Ellipse is 
an invalid shape for locating signal.   True horizontal location of object is on isosigmachron (+).}}

\raggedright
\end{figure}

The points whose sum of {\it signal times} is constant is an isosigmachron, derived from the Greek
words iso for same and sigmachron for sum in time \citep{patent_sbe1}.
This is a  natural geometry for understanding
locations when the effective speeds differ. 
Consider its shape in the ocean where the geometry is identical to the case in 
Sec. IV B2 except the source is at receiver one's position and the receiver is at receiver two's position 
(Fig. \ref{fig:isosigma}).
The measured sum in signal time is 2.7912 + 3.2626 = 6.0538 s. The isosigmachron looks like a distorted ellipse,
and intersects the source's true location. The ellipse is drawn for a sum of distance
given by $c (t_1+t_2) = 1450(6.0538)=8778.0$ m. It does {\it not} contain the true location.  
Confidence limits for location are produced in the same way as described for isodiachrons (Sec. IV B2).

\subsection{\label{subsec:4:3}  Vertical Coordinate Constraint for Object}

We consider how to estimate a reliable confidence interval for an object's location with prior
information of its vertical coordinate. The approach is the same as the previous section except
the grid of points, $(x_k,y_k,z_k)$, is constrained to a smaller subset of the vertical coordinate, $z$.  
For example, suppose we locate a surface ship: we  set $z_k=0$ where the water surface is at zero.
If we know a whale is in the upper 100 m, the grid of points only includes values between zero and 100 m depth.

\subsection{\label{subsec:4:4}  Hybrid 2D and 3D Location Models}

In Sec. IVA, signals are located only when their 2D effective speeds have acceptably small error.
Signal times are discarded when associated errors exceed the threshold.
Instead of discarding these signals, we can estimate horizontal location with a 3D model, and use
its 2D location.  We have to pay the price for computing 3D locations for some but not all the data.

\section{\label{sec:5} Reliable confidence intervals for location accounting for all errors}

The previous section quantified errors in 2D location when the 2D effective speed is
constant.  These errors go to zero if the 2D effective speed is allowed to vary.
Derived locations are typically subject to other errors including  measurements of signal time,
locations of instruments, 3D effective speed  (has no error for electromagnetic signals in a vacuum in the
absence of a gravitational field), and unsynchronized clocks.

We show how to obtain confidence intervals from a 2D model for  
TDOA measured with these errors except clocks are synchronized for simplicity.
Extremely reliable confidence intervals 
are computed with a non-Bayesian method called Sequential Bound Estimation (SBE)
\citep{sbe}.  It solves the non-linear equations for location without approximation.
Analytical solutions for location are obtained with isodiachrons \citep{isodiachrons}:
we allow the 2D effective speed to differ between sections and have uncertainty along each section.
Two sections have probability zero
of having the same 2D effective speed. Because isodiachrons do not extend to infinity, we are guaranteed
locations are finite \citep{isodiachrons} as long as we impose finite bounds on all other variables
affecting location.  The most useful output of SBE is a 100\% confidence interval for location.
To date, this interval always contains the true answer both in tens of thousands of simulations and
experiments having independent measurements for location of the source \citep{sbir_phase2}.
2D effective speed
is constrained to a finite-width interval and simulations include deep and shallow water scenarios
where sound speed varies horizontally and vertically
over a wide variety of bottom profiles. It has been tested independently
by the Navy in deep water \citep{sbir_phase2}. The author used his own commercial software service to run this software, and it
is at Transition Readiness Level 6 \citep{sbir_phase2,trl}.

Inputs to SBE are
100\% intervals for receiver locations, 
2D effective speeds between the source and each receiver, and TDOAs  between each receiver
and receiver number one: all are large enough to contain the true answer.
Since there are five receivers, there are four TDOAs.  Before showing results with SBE,
we derive the horizontal locations where a 2D model is valid.

\subsection{\label{subsec:5:1}  Valid Locations in 2D Models for Sequential Bound Estimation}

Using Eq. (\ref{eq:2d_speed2}), the minimum and maximum 2D effective speeds are,
\begin{equation}
\check{u} = \frac{\hat{c}}{(1+(\hat{v}^2/\check{h}^2)^{1/2}} \ , \label{eq:2d_speed2_min}
\end{equation}
and,
\begin{equation}
\hat{u} = \frac{\check{c}}{(1+(\check{v}^2/\hat{h}^2)^{1/2}} \ , \label{eq:2d_speed2_max}
\end{equation}
respectively.  Bounds for the 3D effective speed, $[\check{c} , \hat{c}]$, are computed by a model or some other method:
they are guaranteed to contain the true 3D effective speed. Similarly, bounds for the vertical distance between
source and any particular receiver are specified with $[\check{v} , \hat{v}]$.  We determine $\check{h}$:  
the minimum horizontal distance of a source from a receiver, by specifying bounds for the 2D
effective speed, $[\check{u} , \hat{u}]$. Invalid regions are the set of points where the horizontal  distance
is less than $\check{h}$.  

We set $\hat{h}$ to be the maximum horizontal distance  of signal detection.
Then we solve for the maximum 2D  effective speed, $\hat{u}$, from Eq. (\ref{eq:2d_speed2_max}) because
all values on its right side are known. We specify the interval width of the 2D effective speeds with,
\begin{equation}
\delta u \equiv \hat{u}-\check{u} = f (\hat{c}-\check{c}) \ , \label{eq:delta_u}
\end{equation}
where the number, $f$, is specified.  Larger values of $f$ are associated with wider bounds.
Reliable 2D location models yield confidence intervals assuming the 2D effective speed is
somewhere within an interval: the larger the interval, the larger the error of location  but the closer
the source can be to a receiver.  This is a natural  trade-off.

To get $\check{h}$, we equate $\delta{u}$ from Eq. (\ref{eq:delta_u}) with $\hat{u}-\check{u}$ from
Eqs. (\ref{eq:2d_speed2_min},\ref{eq:2d_speed2_max}), and solve for the remaining unknown, 
\begin{equation}
\check{h}=\frac{\hat{v}}{\sqrt{a^2-1}} \ , \label{eq:hmin_sbe}
\end{equation}
where,
\begin{equation}
a \equiv \frac{-\check{c}}{\delta u - \frac{\hat{c}}{(1+(\check{v}^2/\hat{h}^2)^{1/2}}} \ .
\end{equation}

\subsection{\label{subsec:5:2}  Example}

\begin{figure}[ht]
\includegraphics[width=\reprintcolumnwidth]{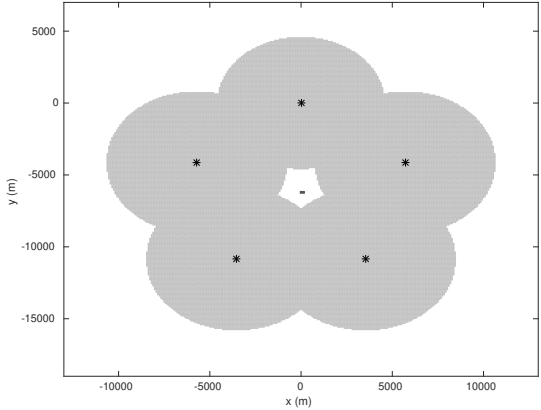}
\caption{\label{fig:invalid_2d_regions}{Extremely reliable 100\% confidence interval for source near center (black) computed from 
TDOAs between signals from five receivers  
(asterisks). Singularities of 2D effective speed at asterisks.
Receiver's depths are a few hundred meters below the source
(Sec. VB). Invalid regions of 2D model are gray. Locations computed with 
sequential bound estimation and isodiachrons \citep{isodiachrons,sbe}.}}

\raggedright
\end{figure}

We assume the acoustic source 
is between 1 and 100 m depth with a maximum horizontal detection range of $\hat{h}$=15 km.
Receivers are a few hundred meters deeper: between 280 and 300 m depth.
They are situated within $\pm 25$ m of the vertices of a pentagon (Fig.  \ref{fig:invalid_2d_regions}). 
The bounds of the 3D effective speed are $\check{c}$=1440 and $\hat{c}$=1455 m/s.  The maximum 2D effective
speed is computed from Eq. (\ref{eq:2d_speed2_max}) using the minimum vertical distance between source
and receiver: $\check{v}$=280-100=180 m: we get $\hat{u}=1454.90$ m/s.  
 
We specify the interval width for 2D effective speed using $f=1.2$  in Eq. (\ref{eq:delta_u}).
The minimum 2D effective speed is $\check{u}$=1436.90 m/s (Eq. (\ref{eq:delta_u})).
Finally, Eq. (\ref{eq:hmin_sbe}) 
yields the minimum horizontal distance of any receiver to the first
valid location: $\check{h}$=4545.9 m,
an astonishingly large  distance for a situation where receivers are only a few hundred meters
deeper than the source and where the receivers are separated by many kilometers. Invalid regions
are shaded gray (Fig. \ref{fig:invalid_2d_regions}). If we wanted valid results
nearer a receiver, we would increase $f$ with attendant increase in the confidence interval for the source's location.

The significance of utilizing reliable confidence intervals is better understood by realizing
$\check{h}=4545.9$ m is {\it not} the same as obtained by solving Eq. (\ref{eq:2d_speed2}) for $h$,
\begin{equation}
h = \frac{v}{\sqrt{(\frac{c}{u})^2-1}} \ , \label{eq:hh}
\end{equation}
and finding its minimum,
\begin{equation}
\check{h} \neq \frac{\check{v}}{\sqrt{(\frac{\hat{c}}{\check{u}})^2-1}} \ , \label{eq:minh_wrong}
\end{equation}
yielding 1130.5 m: smaller than 4545.9 m. The value 1130.5 m is {\it only} true 
if the  vertical  separation is  $\check{v}=180$ m, the maximum 3D effective speed is 1455 m/s, and 
the minimum 2D effective speed is 1436.9 m/s. If we knew these were the only possible values for the
vertical separation  and 3D and 2D effective speeds, $\check{h}=1130.5$ m would be the correct value.
However we do not know the vertical separation, nor the 3D or 2D effective speeds.
Instead we are only certain they fall somewhere within their specified intervals.
Since we require an extremely reliable confidence interval, we enforce their intervals of prior uncertainty,
yielding $\check{h}=4545.9$. 

TDOAs are assumed to be within $\pm 0.02$ s of the true TDOAs. SBE yields a 100\% confidence interval
for the source within the $x$ interval [-66.1, 213] m and the $y$ interval [-6300, -6110] m (small black rectangle,
Fig. \ref{fig:invalid_2d_regions}). These contain
its true location $x=49.1$ m and $y=-6210$ m. In this case, SBE identifies the location of the source
within the valid regime of the 2D model, and yields its reliable confidence interval.
Our reliable location algorithm would not use SBE to compute a reliable confidence interval if the source was
in an invalid region: it would yield a reliable confidence interval with its built-in 3D location model;
the hybrid solution
in Sec.  IIID.

\section{\label{sec:6} Other Flatlands}

Up to this point, we discussed 2D models with planar coordinates.
Sometimes, horizontal coordinates are desired in
latitude and longitude, and the 2D model surface is a sphere or spheroid \citep{lewis}.
We discuss the 2D effective speed for the sphere because it is simpler.

\begin{figure}[ht]
\includegraphics[width=\reprintcolumnwidth]{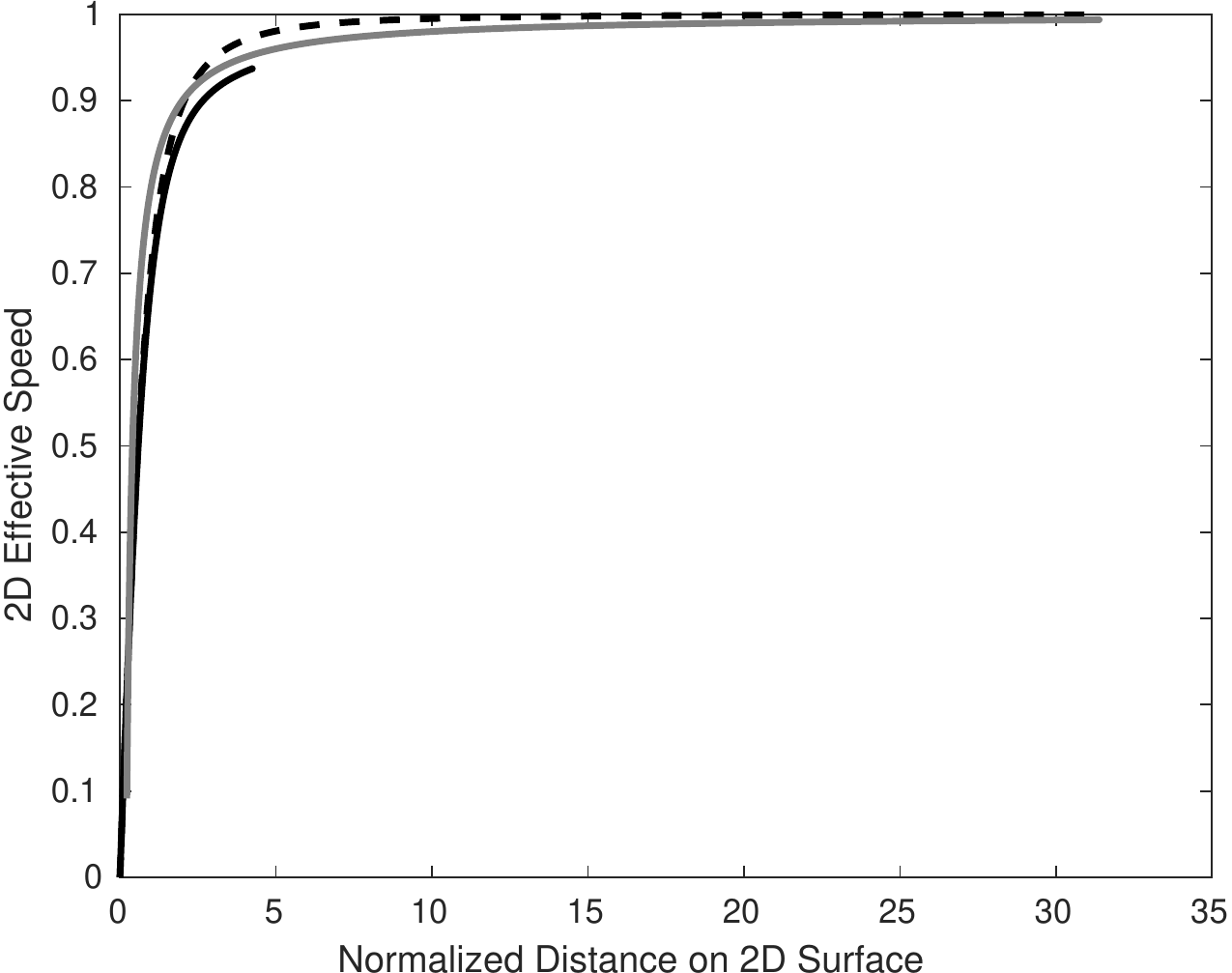}
\caption{\label{fig:eff_speeds_plane_sphere}{2D effective speeds for  planar (dashed) and spherical (solid black) 2D models.
Planar 2D effective speeds computed with Eq. (\ref{eq:2d_speed2}) for 3D effective speed $c=1$, vertical offset, $v=0.1$,
and horizontal axis showing horizontal distance, $h$, from instrument divided by $v$. Spherical 2D effective speeds
computed with Eq. (\ref{eq:u_sphere}) (Sec. VI). It is possible to find an infinite number of coordinate systems
in general relativity where the radial speed of light with respect to a black hole equals values for
planar and spherical 2D effective speeds
when the local speed of light is defined to equal one  (Eqs. \ref{eq:R},\ref{eq:integrate_equate_c}). 
Gray curve is functional form for the radial speed of light in Schwarzschild's coordinates
for an event horizon of radius $0.01$ and a local speed of light equal to one (Sec. VII).}}
\raggedright
\end{figure}

Assume the sphere has radius, $R$, and the object to locate is above the sphere at radius $\rho=R+v$ with $v\geq0$.
As before, assume the 3D effective signal speed is $c$. In the context of the 2D model, signals 
propagate along great circles of length $h$ on the 2D spherical surface, instead of straight
line segments of planar 2D models (Sec. II). The Euclidean distance between the object
and instrument is $d=[(\rho \sin \theta)^2 + (R-\rho \cos \theta)^2]^{1/2}$, where $\theta$ is the angle between
two line segments, the first between the sphere's center and the instrument, and the second between
the sphere's center and the object. The horizontal
distance of the signal path on the sphere is $h=R\theta$. The 2D effective speed for the sphere is obtained
by substituting this $h$ and $d$ into Eq. (\ref{eq:2d_speed1}),
\begin{equation}
u_{sphere} = \frac{cR\theta}{\bigl[2R^2(1-\cos \theta) + 2vR(1-\cos \theta) + v^2  \bigr]^{1/2}} \ . \label{eq:u_sphere}
\end{equation}
Although this 2D effective speed goes to zero when $h$ goes to zero, its functional form is not the same
as the planar 2D effective speed (Eq. \ref{eq:2d_speed2}).
For small horizontal separation, ($\cos \theta ~\sim 1)$, we get $u_{sphere} \sim ch/v$: the same form 
as the  planar model to leading order in $h$ (Eq. \ref{eq:2d_speed2}). If we assume the direct path does not propagate
through the spherical surface, a signal is received when $|\theta| \leq \cos^{-1} {R/\rho}$, or $h \leq R \cos^{-1} {R/\rho}$.
The planar and spherical 2D effective speeds differ (Fig. \ref{fig:eff_speeds_plane_sphere}).
Spherical 2D effective speeds do not exist when  $h> R \cos^{-1}{R/\rho} = 0.43$. Since the figure shows
normalized horizontal separation, $h/v$, values do not exist  when $h/v > 0.43 / 0.1 = 4.3$. 

When the 2D model surface is a spheroid \citep{lewis}, there is no closed-form solution for the 2D effective speed
because geodesic length, $h$, does not have a closed-form expression. Instead,
2D effective speeds are computed with Eq. (\ref{eq:2d_speed2}) and $h$ is computed numerically.

\section{\label{sec:7} Summary and Conclusions}

Signal times have been used for a century to locate signals on 2D surfaces even though objects
are not usually on this surface. Their ubiquitous use up to the present
suggests correctness of approach \citep{bateman,ishmael,sbe,spermwhale,warner_and_dosso,cell_phones,jin,sbe}.
The apparent discovery here of signal-speed singularities in 2D models suggests
findings in thousands of papers could be re-evaluated.  Many results must be approximately correct, while others must not be.
We presented one approach to quantify the validity of 2D models and means to compute extremely reliable
confidence intervals for location (Sec. V). The traditional 2D model with constant speed is most accurate at distances
far from the instruments: the further the better, because the 2D effective speed approaches the correct 3D effective speed
(Fig. \ref{fig:eff_speeds_plane_sphere}). 2D models constraining the speed between finite positive-valued bounds are also
most accurate at distances far from the instruments when those bounds contain the bounds for the 3D effective speed.
This is why invalid regions of these 2D models are close to the instruments, and the valid regions occur
far away (Fig. \ref{fig:invalid_2d_regions}).  The method for determining where 2D models are valid is applicable
at any distance, not just those treated up to 15 km from the instruments (Fig. \ref{fig:invalid_2d_regions}).

2D effective speeds are zero at the horizontal coordinates of
the receivers and monotonically increase to 3D effective speed at infinite distance (Eq. \ref{eq:2d_speed2}).
In a 3D coordinate system, effective speeds do not exhibit singularities.
Locations in 2D can be interpreted geometrically.  For
direct-path times, geometry is conventional. Locations are visualized by intersecting circles: the projection of a sphere
on a flat surface.  For TDOA, 2D effective speeds
can differ by large factors between the object and each receiver. Location is visualized by
intersecting isodiachrons \citep{isodiachrons}: the replacement for hyperbolas 
when the propagation speed of the signal is spatially inhomogeneous.  For bistatic times, location can
be visualized by intersecting isosigmachrons
instead of ellipses \citep{patent_sbe1}.
Collapsing  a 3D problem onto 2D breaks the symmetry of speed.
Geometries are transformed from shapes invented by ancient
Greek mathematicians into geometries  of the modern age wherein locations are derived
with signal times and spatially inhomogeneous speeds \citep{isodiachrons,patent_sbe1}. 

Traditional 2D models yield large errors 
near the instruments.  Of course an experimentalist places instruments
near signals of interest, thinking errors will be smaller. 
This is exactly where singularities occur, leading
to large errors with traditional methods and small errors with  non-traditional methods (Sec. V).

Perhaps the most problematic issue is the common use of TDOA and hyperbolas to locate signals with 2D  models.
When the {\it only} error is due to the use of hyperbolas, locations can be incorrect
by many factors of the vertical separation between the object and the receivers 
(Table I, Fig. \ref{fig:model_dist_errors1_4rec_tdoa}). Error is caused by hyperbolas whose
use is predicated on the assumption of homogeneous speeds. Since 2D effective speeds
often vary by large factors, and since hyperbolas extend to infinity, errors are large.
Even if data were pre-scanned to eliminate locations yielding large errors for a specific set of
receiver coordinates, other nearby receiver coordinates, within tolerance, would generate other locations
with large error. This mess can be eliminated by working with sequential bound estimation and 
isodiachrons \citep{sbe,patent_sbe1}.

The problems we discuss are fundamentally subjects of coordinate systems.
Physics cannot depend on coordinates and correct location and its confidence
must be independent of coordinate-frame. Almost all previous 2D models set
signal speed to a constant, independent of an object's location. In a small number of cases, 2D models
constrain signal speeds to an interval of finite-width \cite{sbe,warner_and_dosso}, and the reported interval does not ever
appear to include zero.  The surfaces utilized by 2D models must allow speed to vary
with location to yield correct location. This fact appears to be new to science.  Recently-invented
geometrical shapes, isodiachrons and isosigmachrons, emerge as a natural way to interpret location.

In summary, signal speed depends on location in one coordinate system
but not another; signal speed singularities appear in one coordinate system but not another,
the geometry for obtaining and interpreting location depends on the coordinate system, and
physics is simpler in one coordinate system (3D) than another (2D).

This is the same behavior
as light described by the physics of general relativity and black holes, and by the unification
of electromagnetism and general relativity.
General relativity assumes the speed of light is constant in a local
coordinate system, even near a black hole. However,  in Schwarzschild's coordinates, speed varies
with location as
\begin{equation}
  dr/dt =(1 - r_s/r ) c \ , \label{eq:schwarz}
\end{equation}
where  $t$ is time measured by a clock at
infinite distance from the black hole, $r$ is zero
at its center, $2\pi r_s$ is the circumference of a circle on the event horizon, and
$c= 299,792,458$ m/s is the speed of light in local coordinates \citep{light_blackhole}.
The decrease in light speed in a gravitational field is experimentally verified,
and is known as the Shapiro effect \citep{shapiro}.
In local coordinates,
there is no singularity but there is in Schwarzschild's coordinates, where light speed is zero
at the event horizon.  In local coordinates, space-time is flat, but in Schwarzschild's 
and all non-local coordinate systems, space-time is described by Riemann metrics and  geometries \citep{gravityzee}.
Regarding unification, in 1919, Kaluza sent Einstein a paper showing how to unify electromagnetism and gravity
by adding a fifth dimension to the four of general relativity, space and time \citep{kaluza,gravityzee}.
The fifth dimension made it easier to understand these otherwise disparate fields \citep{gravityzee}.

The similarity between general relativity and 2D models becomes even closer realizing
it is possible to choose another coordinate system
where the radial speed of light has the same form as the planar 2D model, namely
Eq. (\ref{eq:2d_speed2}). This other coordinate system cannot be
Schwarzschild's coordinates.  Its derivation is due to Dr. J. Khoury at U. Pennsylvania.
We wish to find a new radial coordinate, $R=R(r)$ such that,
\begin{equation}
dR/dt = c/(1+(a/R)^2)^{1/2} \ , \label{eq:R}
\end{equation}
and $R$ is interpreted as a proper distance in this to-be-determined other coordinate frame.
We can and will assume time is measured by the same $t$ as in Schwarzschild's coordinates, so
this other frame's proper time is still $t$.
The proper distance is the ruler used by an observer in the other frame. Equating $c$ from Eqs (\ref{eq:schwarz})
and (\ref{eq:R}),
\begin{equation}
dR (1+(a/R)^2)^{1/2} = dr/(1-r_s/r) \ . \label{eq:equate_c}
\end{equation}
Integrating both sides, we get,
\begin{eqnarray}
  (a^2 + R^2)^{1/2} &-& \frac{a}{2}\log \biggl(\frac{(a^2+R^2)^{1/2} + a}{(a^2+R^2)^{1/2} - a}  \biggr) \nonumber \\
  &=& r_s \log(r-r_s) + r \ , \label{eq:integrate_equate_c}
\end{eqnarray}
where we set the additive constant to zero for simplicity.
This yields an implicit relation to find the ruler, $R(r)$, needed to yield the observed radial speed of light
in Eq. (\ref{eq:R}). From Eq. (\ref{eq:integrate_equate_c}), when $R \to 0$, the left side
of Eq. (\ref{eq:integrate_equate_c}) goes to negative infinity.  The right side must also go to negative
infinity which occurs only when $r=r_s$. Therefore, $R=0$ at Schwarzschild's event horizon, $r_s$.
On the other hand, when $r \to \infty$, the right side is dominated by $r$ and the left side is dominated
by $(a^2+R^2)^{1/2} \to R$ when $R$ is large. So when $r \to \infty$, $R \to \infty$.
Using this same procedure, it is possible to find yet
another coordinate system in general relativity yielding a radial speed of light with the same form as a spherical 2D model
(Eq. \ref{eq:u_sphere}). 

In general relativity, there are an infinite number of other coordinate systems yielding
the same form as Eq. (\ref{eq:2d_speed2}) if we allow variation in both proper time and proper length,
instead of just proper length as above. These are problems of coordinate systems.

Even though an exact match is possible, values for the radial speed of light in Schwarzschild coordinates are similar to the values
for the 2D effective speed from planar and spherical geometries (Fig. \ref{fig:eff_speeds_plane_sphere}).
In this figure, we change units so the speed of light equals one far from a black hole (or one in local
coordinates).  

In light of the common role of coordinate systems and metrics in understanding general relativity, black holes, and 2D and 3D
effective speeds,
it makes sense to call singularities in 2D models ``2D black holes.''
The radius of their event horizon is zero.  Similarly, we can refer to invalid regions of
2D models as ``2D shadows.''.  They always contain one or more 2D black holes.
2D black holes are not the same as sonic black holes, a phenomenon
predicted by Unruh in 1981, where sound has difficulty escaping from a current exceeding the local 
speed of sound \citep{unruh}.


\begin{acknowledgments}
  I thank Professor Justin Khoury at the U. Pennsylvania for discussions concerning gravitational black holes,
  and his derivation of a coordinate system yielding the same form for the radial speed of light
  as for planar 2D effective speed.
  I thank Professors Robert Giegengack and Mary Putt (U. Pennsylvania) for their insightful comments. The editor
  and reviewers provided comments leading to clearer presentation of the material. Dr. Spiesberger notes a conflict
  in interest in citing the availability of his commercial software service for locating signals with SBE. 
Time for composing this paper was provided by ONR contract N0001417C0230.
\end{acknowledgments}







\end{document}